\documentclass[12pt,preprint]{aastex}
\usepackage{natbib}
\usepackage{epsfig}
\usepackage{graphicx}
\usepackage{amssymb}
\usepackage{multirow}
\usepackage{longtable}

\newcommand{\msun}{\rm \,M_{\odot}}

\begin{document}

\author 
{J. Boyles$^{\dagger,1,2}$, R. S. Lynch$^{3,4}$, S. M. Ransom$^{5}$, 
I. H. Stairs$^{6}$, D. R. Lorimer$^{1,7}$, M. A. McLaughlin$^{1}$, 
J. W. T. Hessels$^{8,9}$, V. M. Kaspi$^3$, V. I. Kondratiev$^{8,10}$, 
A. Archibald$^3$, A. Berndsen$^6$, R. F. Cardoso$^1$, A. Cherry$^6$, 
C. R. Epstein$^{11}$, C. Karako-Argaman$^3$,
C. A. McPhee$^6$, T. Pennucci$^4$, M. S. E. Roberts$^{12}$, 
K. Stovall$^{13,14}$, \& J. van Leeuwen$^{8,9}$ \\
$^{\dagger}$ Enquiries to:jason.boyles@wku.edu \\
$^1$ Department of Physics, West Virginia University, Morgantown, WV 26506 USA  \\
$^2$ Department of Physics and Astronomy, Western Kentucky University, Bowling Green, KY 42101 USA \\
$^3$ Department of Physics, McGill University, 3600 University St., Montr\`{e}al, Quebec, H3A 2T8, Canada \\
$^4$ Department of Astronomy, University of Virginia, Charlottesville, VA 22904-4325, USA \\
$^{5}$ National Radio Astronomy Observatory (NRAO), 520 Edgemont Road, Charlottesville, VA 22901, USA \\
$^6$ Department of Physics and Astronomy, University of British Columbia, 6224 Agricultural Road, Vancouver, British Columbia V6T 1Z1, Canada \\
$^7$ Also adjunct at National Radio Astronomy Observatory, Green Bank, WV 24944, USA \\
$^8$ ASTRON, the Netherlands Institute for Radio Astronomy, Postbus 2, 7990 AA, Dwingeloo, The Netherlands \\
$^9$ Astronomical Institute "Anton Pannekoek," University of Amsterdam, Science Park 904, 1098 XH Amsterdam, The Netherlands \\
$^{10}$ Astro Space Center of the Lebedev Physical Institute, Profsoyuznaya str. 84/32, Moscow 117997, Russia \\
$^{11}$ Department of Astronomy, Ohio State University, 140 W. 18th Avenue, Columbus, OH 43210 \\
$^{12}$ Eureka Scientific, 2452 Delmer Street, Suite 100, Oakland, CA 94602-3017, USA  \\ 
$^{13}$ Center for Advanced Radio Astronomy and Department of Physics and Astronomy, University of Texas at Brownsville, Brownsville, Texas 78520 USA  \\ 
$^{14}$ Department of Physics and Astronomy, University of Texas at San Antonio, San Antonio, Texas 78249 \\ }


\title 
{The Green Bank Telescope 350 MHz Drift-scan Survey I: Survey
Observations and the Discovery of 13 Pulsars}

\maketitle

\section*{Abstract}
\label{abs}
Over the summer of 2007, we obtained 1191 hours of `drift-scan' 
pulsar search observations with the Green Bank Telescope at a radio frequency of 350 MHz.  
Here we describe the survey setup, search procedure, and the discovery and follow-up timing of 
thirteen pulsars. Among the new discoveries, one (PSR J1623$-$0841) was discovered 
only through its single pulses, two (PSRs~J1327$-$0755 and J1737$-$0814) are millisecond 
pulsars, and another (PSR~J2222$-$0137) is a mildly recycled pulsar.
PSR J1327$-$0755 is a 2.7 ms pulsar at a DM of 27.9~pc $\rm cm^{-3}$ in a 8.7 day orbit with 
a minimum companion mass of 0.22$\msun$.  
PSR~J1737$-$0814 is a 4.2 ms pulsar at a DM of 
55.3~pc $\rm cm^{-3}$ in a 79.3 day orbit with a minimum companion mass of 0.06$\msun$.  
PSR J2222$-$0137 is a 32.8 ms pulsar at a very low DM of 3.27~pc $\rm cm^{-3}$ in a 2.4 day orbit with
a minimum companion mass of 1.11$\msun$. It is most likely a white dwarf-neutron star system or 
an unusual low-eccentricity double neutron star system.  Ten other pulsars discovered in this survey 
are reported in the companion paper Lynch et al. 2012.  

\begin{keywords}
   K{\bf Key Words:} stars:neutron -- pulsars:general -- pulsars:surveys -- pulsars:individual (PSR~J1327$-$0755, PSR~J1623$-$0841, PSR~J1737$-$0814, PSR~J1941$-$0121, PSR~J2222$-$0137) 
\end{keywords}

\section{Introduction}
\label{intro}

Since the discovery of the first pulsar in 1967 (Hewish et al. 1968),
many fascinating sources and experiments have been associated with pulsars.   
These include but are not limited to tests of general relativity (e.g. Lyne et al. 2004, Kramer et al. 2006), studies of
binary evolution (e.g. Bhattacharya \& van den Heuvel 1991), probes of electron density models of the interstellar medium
(e.g. Cordes \& Lazio 2002), constraints on the neutron star equation of state
(e.g. Lattimer \& Prakash 2001, Lattimer \& Prakash 2007), and use in a pulsar timing array  
for gravitational wave detection (e.g. Hobbs et al. 2010, Demorest et al. 2012).

Searches for pulsars are very demanding, requiring both significant amounts of
computational and human resources.  The high time and frequency resolution
needed for pulsar searches result in data rates greater than for most astronomical
projects (typically 100 GB/hr or more).  Large computing clusters are needed to
thoroughly search for both periodic and transient dispersed signals.
After all the computer searches are completed, candidates are typically visually inspected
to determine likely pulsar candidates, though automated algorithms 
are becoming more widely used for sifting through large  
numbers of candidate signals (e.g., Eatough et al. 2010).

Pulsars are brighter at low frequencies (i.e. a few hundred MHz) but pulsar survey sensitivity is reduced 
by dispersion smearing, higher sky temperature, and scattering at these frequencies.  Therefore most 
recent large-scale pulsar surveys like the Parkes Multibeam Pulsar Survey (PMPS; Manchester et al. 2001), 
High Time Resolution Universe Pulsar Survey (HTRU; Keith et al. 2010), and the Pulsar Arecibo
L-band Feed Array Survey (PALFA; Cordes et al. 2006)
are conducted around 1.4 GHz.  Lower frequency observations
($<$ 500 MHz) are better used to locate weak sources away from the Galactic plane where
the effects of scattering, higher sky temperature, and dispersion broadening are minimized. 

The National Radio Astronomy Observatory (NRAO) in Green Bank, WV has a long history of
successful searches for radio pulsars at low frequencies.
Damashek et al.~(1978) discovered 17 new pulsars with the 92-meter (300$^\prime$) transit
telescope at a frequency of 400 MHz.   Later, using the same telescope at a
frequency of 390 MHz, Dewey et al.~(1985) found 34 new long-period
pulsars.  However, they were not sensitive to newly discovered millisecond pulsars (MSPs; P $<$
20 ms; Backer et al. 1982).  The first low-frequency
 survey to be sensitive to MSPs was conducted by Stokes et al.~(1985) at 390 MHz.
While this survey did not discover any MSPs, it did find 20 new long-period pulsars.
Sayer et al.~(1997) used the 43-meter (140$^\prime$) telescope at Green Bank at a frequency of
370 MHz and discovered eight new pulsars including one mildly relativistic binary pulsar.  
New backend hardware and higher data rates now allow frequency resolution that was previously
unattainable by earlier observing systems.  Most recently, a North Galactic
plane survey was conducted with the Robert C. Byrd
Green Bank Telescope\footnote{The Robert C. Byrd Green Bank
Telescope (GBT) is operated by the National Radio Astronomy Observatory.}   
(GBT) by Hessels et al.~(2008) at 350 MHz
and discovered 33 new pulsars.  None of these were MSPs despite the survey's sensitivity to these 
systems\footnote{The North Galactic plane survey has only been partially 
processed to look for MSPs.}.

Our survey took advantage of a unique opportunity when the GBT was
immobilized for track refurbishing during the northern summer of 2007. During this time,
we recorded data at a radio frequency of 350 MHz as the sky
passed through the beam of the telescope.
We will discuss these drift-scan observations in Section~2.
In Section~3 we describe the follow-up observations and timing
analysis of the new pulsars.  In Section~4 we present the properties of the thirteen pulsars and
in Section~5 we discuss interesting individual
objects.  Finally in Section~6 we present the conclusions and the future
survey planned in light of the results obtained in this survey.  The survey
sensitivity, pipeline, and the remaining new pulsars are discussed
in Lynch et al. (2012; Paper II).  

\section{Survey Observations}\label{obs}

The drift-scan observations occurred from May through August in 2007.
The telescope was parked at a set azimuth during the
track refurbishing and data were recorded while the sky drifted through the
telescope's beam.  The elevation of the telescope was adjusted and had a limit 
of $>$ 60$^\circ$ during the weekdays and $>$ 30$^\circ$ during the weekends.  
The azimuth was set at $\sim$229$^\circ$ for the first
half of the observations and $\sim$192$^\circ$ for the second half of the
observations.  These azimuth restrictions allowed us to access declination ($\delta$) 
ranges of $-$7.72$^\circ$ $< \delta <$ 38.43$^\circ$ and $-$20.66$^\circ$ $< \delta <$
38.43$^\circ$ for azimuths of 229$^\circ$ and 192$^\circ$, respectively.  All
right ascensions were available for us to observe.

We chose a frequency of 350 MHz for this survey for many reasons.
Firstly, at 350 MHz, the radio frequency interference (RFI) environment at the GBT is
well known and remarkably manageable.  Secondly, pulsars generally have very steep spectral indices (e. g. Sieber 1973)
and our survey covers Galactic latitudes away from the Galactic plane, where propagation effects 
in the interstellar medium are generally less severe.  
Lastly, the GBT's beam size at 350 MHz is $\sim$36$\arcmin$, allowing us to cover a
greater amount of the sky than would be possible with higher frequency receivers.
The sky coverage rate (210 deg$^2$ per day) is faster than that achieved by the
PMPS at 1.4 GHz (26 deg$^2$ per day; Manchester et al. 2001).

The now-retired Spigot autocorrelation spectrometer (Kaplan et al. 2005) was used for the
GBT drift-scan observations.  We used
50 MHz of bandwidth with 81.92-$\mu$s sampling time.  For the first
week of observations (MJDs 54222 -- 54230) data were recorded with 1024 three-level
auto-correlation functions which were accumulated into 16-bit registers for analysis.
After that time, and until the conclusion of the observations, a new 2048 three-level auto-correlation mode was 
used which accumulated data into 8-bit registers.

\begin{figure}[t]
\includegraphics[scale=0.6,angle=270]{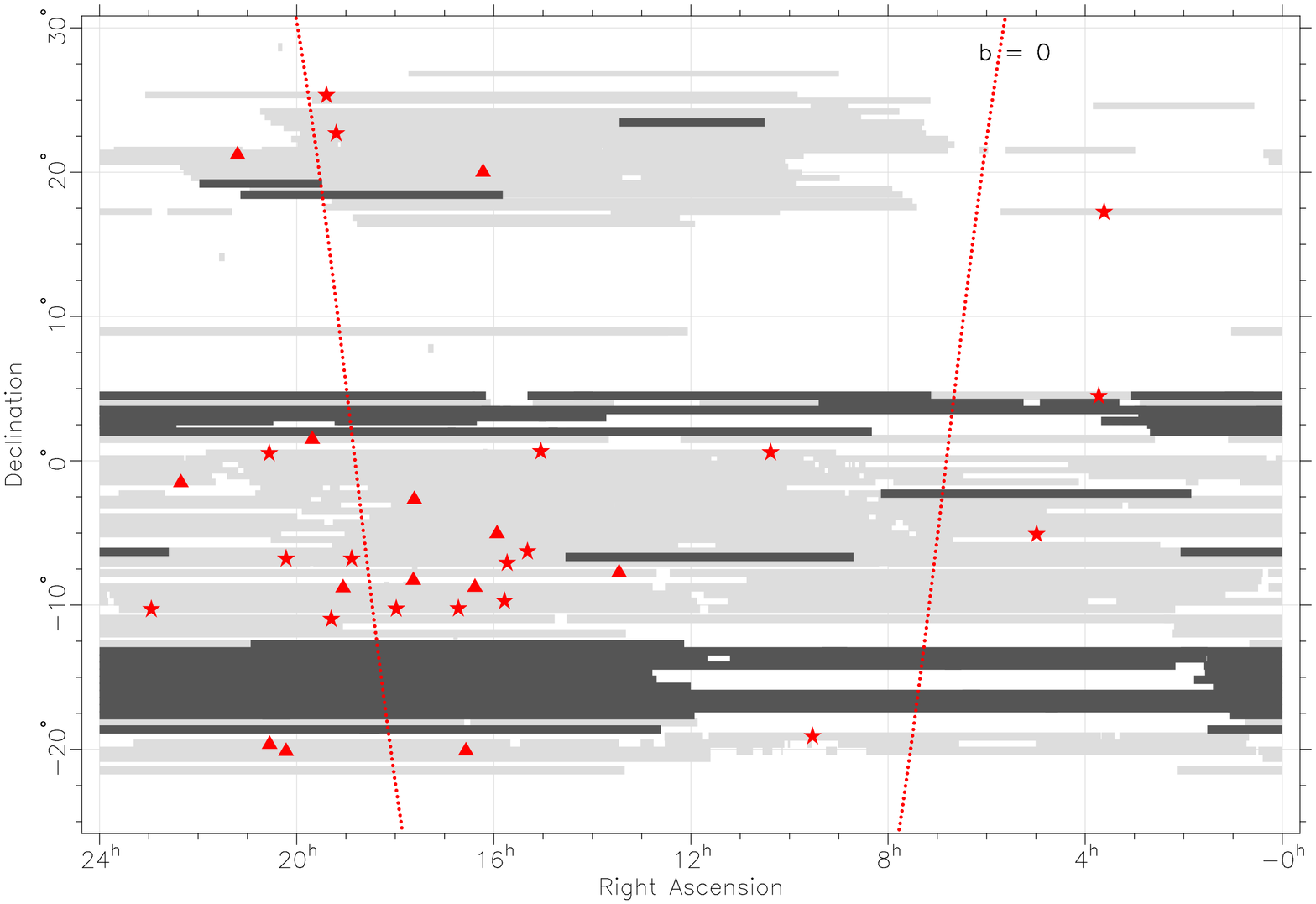}
\caption{\label{fg:cov}
The sky coverage of the survey shown in Right Ascension and Declination (J2000 coordinates). The Pulsar 
Search Collaboratory 
portion is shown in dark gray and the red dotted line shows the Galactic plane.  New discoveries presented 
in this paper are shown as filled red triangles, while all other GBT drift-scan discoveries are shown as 
filled red stars.}
\end{figure}

Data were recorded every night on each weekday and for 24 hours a day during the
weekends.   In total, 1491 hours of data were taken, amounting to 134 TB
(see Supplemental Online Material for the observation list and Figure~\ref{fg:cov} for sky coverage area).  
This amounts to a sky coverage of 10347 square degrees, of which $\sim$2800 went
to the Pulsar Search Collaboratory (PSC; Rosen et al. 2010; Rosen et al. 2012).\footnote{The PSC is a
collaboration between NRAO and WVU that involves high-school students in searching for pulsars.
http://www.pulsarsearchcollaboratory.com/}
A source passing through the center of the beam is visible for roughly 140 s.  Assuming
a detection significance of 8$\sigma$, a digitization correction factor of 1.16 to account for 3-level quantization  
of the signal, and a pulse duty cycle of 0.04 results in a survey flux density limit
\vskip -0.45cm

\begin{equation}
S_{\rm min} = 0.012 T_{\rm sys} \frac{\rm mJy}{\rm K},
\label{eq:eqrad}
\end{equation}

\vskip -0.20cm
\noindent where $T_{sys}$ is the total system temperature.  The system temperature has two major 
components: the receiver temperature, which is $\sim$22 K, and the sky temperature.  The sky temperature
can vary greatly in the survey region.  Far off the plane,  the sky temperature is $\sim$30 K at 350 MHz
but near the plane the sky temperature can be 200$-$400 K in our survey regions.  For a total system 
temperature of 52 K, a survey sensitivity limit of 0.6 mJy is obtained.  
Scaling our flux density limit with an average pulsar spectral index of $-$1.8
(Maron et al. 2000) gives us a limit of $\sim$0.05 mJy at 1352 MHz, more sensitive than the mid- and high-latitude
HTRU survey in the same regions, but our observations were hampered more significantly by dispersion 
smearing and scattering.  A more detailed discussion of the survey sensitivity 
is presented in Paper II.  Data processing was done with the
\texttt{PRESTO}\footnote{http://www.cv.nrao.edu/$\sim$sransom/presto/}
software package (Ransom 2001) and the data pipeline will be presented in Paper II.

\section{Timing Observations and Analysis}\label{tim}

Timing observations began in June of 2008.  At that stage,  
nine new pulsars that had been discovered were observed.  The first session was 
conducted at a central frequency of 350 MHz with the GBT using the Spigot backend with 50 MHz of bandwidth 
and 81.92-$\mu$s sampling time.  With observations spread across six hours, this session was 
used to establish a very accurate period and for gridding of pulsars to find a more accurate 
position (Morris et al. 2002).  

Most of the rest of the follow-up timing observations in the first year (July 2008 
$-$ May 2009) were conducted using the Spigot backend at a central frequency of 820 MHz  
with 1024 channels, 50 MHz of bandwidth and 81.92 $\mu$s sampling time.  All further 
observations (June 2009 onwards) were conducted using the Green Bank Ultimate 
Pulsar Processing Instrument (GUPPI) backend 
(Ransom et al. 2009).  Most of the observations used a center frequency of 
820 MHz with 200 MHz of bandwidth, 2048 channels, and 40.96 $\mu$s sampling time.  
For pulsars that had satisfactory timing solutions, the data were taken in on-line 
folding mode with ten-second integrations, while the rest were done in search mode, 
with the data recorded in {\texttt{PSRFITS}} format (Hotan et al. 2004).

Each search-mode data set was first subbanded, or de-dispersed at the pulsar's dispersion measure (DM) 
into broader frequency channels, using \texttt{PRESTO}. We used 16 
subbands for long-period pulsars and 32 subbands for MSPs.  
The DM value used for subbanding is either the discovery value or a timing-derived value,
if multi-frequency times of arrival (TOAs) had allowed fitting the DM.
The on-line folded GUPPI data require no subbanding.

\begin{figure}
\includegraphics[scale=0.6,angle=270]{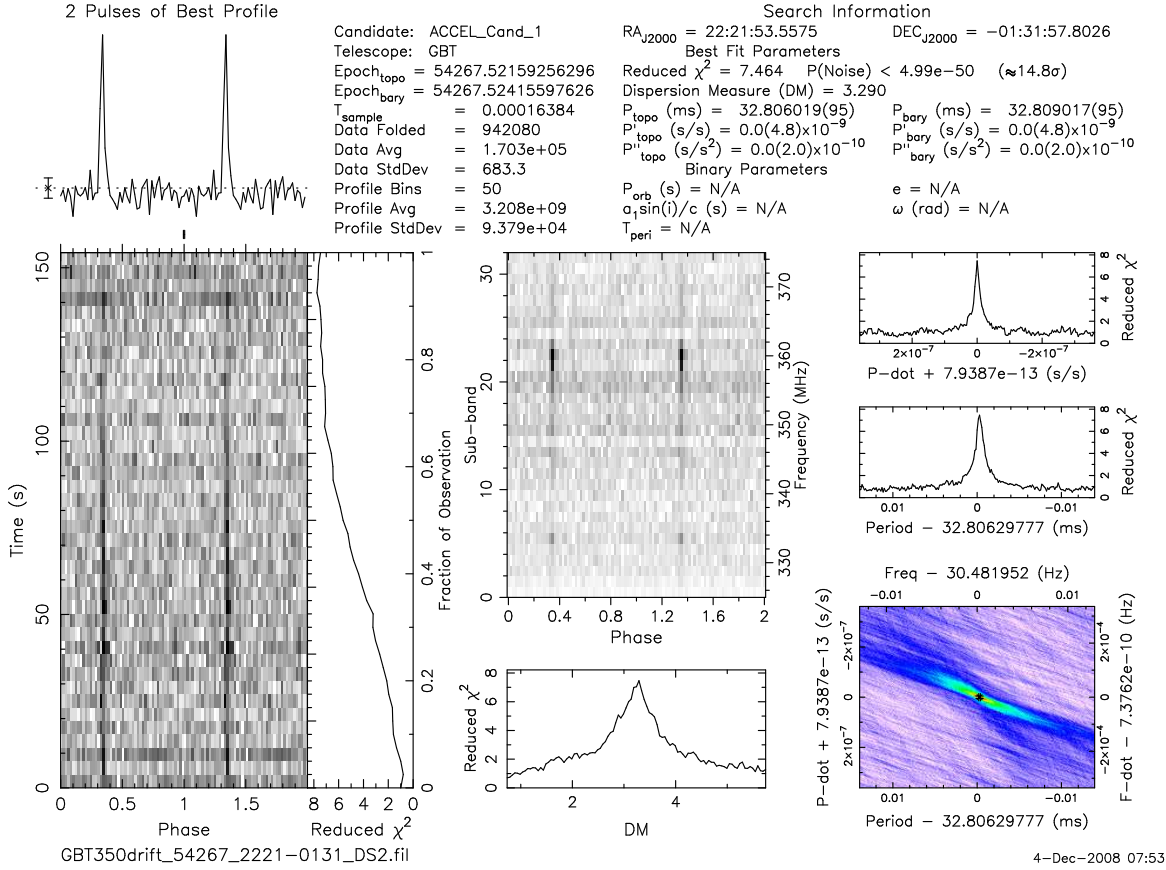}
\caption{\label{fg:ex}
Example of the search output diagnostic plot for new discovery PSR J2222$-$0137. Left--hand side 
shows a pulse profile at the top and a plot of intensity versus time on the bottom.  
Along the plot of intensity versus time is a plot of accumulated $\chi^2$ versus time.  The middle 
top plot shows observing frequency versus pulse phase and the bottom shows $\chi^2$ versus DM.  
The right hand--side (from top to bottom) shows $\chi^2$ versus period, $\chi^2$ versus 
$\dot{P}$, and a $\chi^2$ intensity plot of period-$\dot{P}$ space.  }
\end{figure}

Folded pulsar profiles were created from the subbanded data using \texttt{PRESTO}; an example output
plot is shown in Figure~\ref{fg:ex}.  These folds were done using the best available ephemeris
with 128 bins for normal period pulsars or 64 bins for MSPs.  On-line folded GUPPI data were
co-added in time and frequency and flux calibrated (described later) to form a profile using the
\texttt{PSRCHIVE}\footnote{http://psrchive.sourceforge.net/current} package with 256
pulse phase bins.

TOAs were measured from the folded profiles using the frequency-domain algorithm in \texttt{PRESTO}
for all search-mode data (Taylor 1992).  We measure 
one TOA per observation for each isolated pulsar and three per
observation for each binary pulsar for better determination of binary parameters.  TOAs for the 
on-line folded data were created with \texttt{PSRCHIVE} and also used the Taylor (1992) method with
the same number of TOAs as the search-mode data.  Templates were created by fitting multiple Gaussian functions
to the summed pulse profile for each frequency and backend.  From these Gaussian components, 
noise-free templates were created for each observing frequency with the phase of the fundamental component 
in the frequency domain rotated to zero.  

For each pulsar, the TOAs and ephemeris files were used in \texttt{TEMPO} to carry out a standard pulsar
timing analysis (see Lorimer \& Kramer 2005 and references therein).  The Solar System model DE405 
(Standish 1998) and the time standard TT(BIPM) were used in \texttt{TEMPO}.  
Time delays due to using two different backends were taken into account by placing jumps around 
the Spigot data, allowing a possible time offset due to each individual instrument.  
The result was a model for the behavior of the pulsar.  The goodness of the fit was quantified by the  
root-mean-square of the difference between the model and the TOAs.  The timing-derived parameters for 
each pulsar are listed in Tables 2 and 3 with doubled 1$\sigma$ errors listed to provide a conservative 
estimate of the 68$\%$ confidence limit.  The root-mean-square 
values are on the order of a few milliperiods.  Since no systematic trends are seen in the residuals, 
this is interpreted as an underestimation of the individual TOA errors.  As is a common practice in 
pulsar timing, the TOA uncertainties are scaled by a constant factor, EFAC, to bring the reduced $\chi^2$ 
of each timing model to unity.  

\section{Other Measured Parameters}\label{non}

\subsection{820-MHz Fluxes}\label{flux}
Fluxes were measured using the calibration routine associated with \texttt{PSRCHIVE}
with the on-line folded GUPPI data.  On- and off-source scans of the extra-galactic radio source 3C190 
were used for the flux calibration.  Before each observation, a one-minute calibration scan was taken 
with a 25-Hz noise diode at the receiver.  RFI was removed from the pulsar and calibration scan data 
using the {\tt psrzap} utility from the \texttt{PSRCHIVE} package.  

\texttt{PSRCHIVE} calibration was applied to the data using on- and off-flux scans, calibration scan, 
and pulsar data that allowed slightly different levels in the two polarizations (Hotan et al. 2004).
These files were used to calculate an absolute flux conversion factor which  was applied to the
pulsar data.  Lastly, a flux measurement was taken from the calibrated data.

\subsection{Rotation Measure}\label{rm}

For pulsars with enough linearly polarized flux, rotation measures (RMs) were 
calculated. This includes 10 of the 13 pulsars in this sample.  \texttt{PSRCHIVE} was used with 
on-line folded GUPPI data for producing the RMs.  The method 
used here was similar to that presented in the recent work of Yan et al.~(2011).

\begin{figure}
\includegraphics[scale=0.70,angle=0]{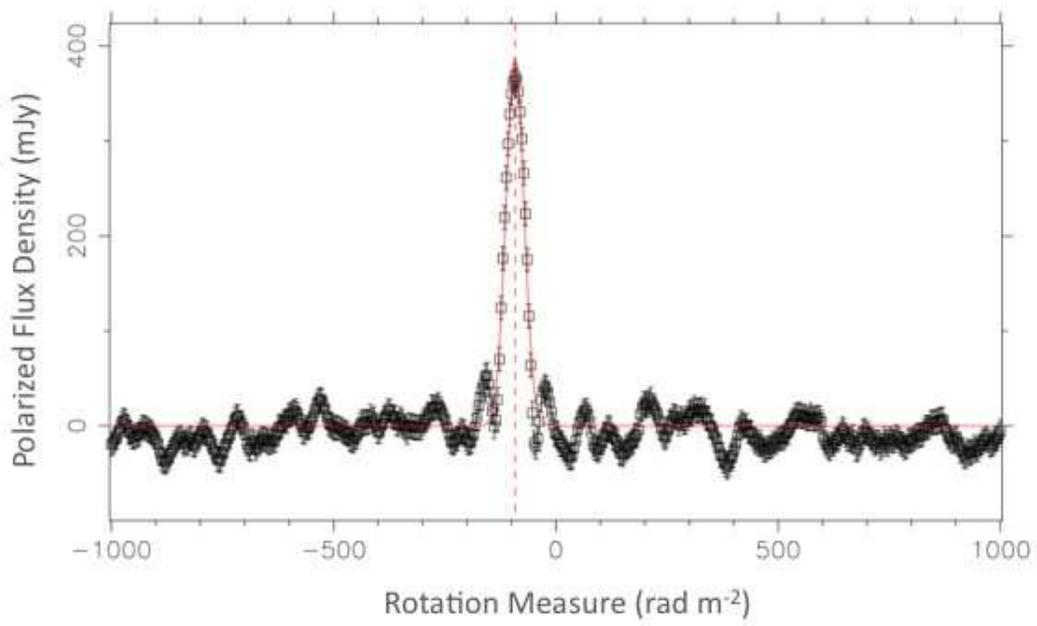}
\caption{\label{fg:rmsamp}
Example fit for RM for PSR J1941+0121.  The data points are square boxes with error bars and the best-fit RM is
shown by a vertical dotted line.}
\end{figure}

After the data were calibrated, we tested many
different RM values from $-$1000 to 1000 $\rm rad~m^{-2}$ to find the one which provided the most
polarized flux.  An example plot showing the results of this procedure for PSR J1941+0121 can be seen in
Figure~\ref{fg:rmsamp}.  A Gaussian was fit to the polarized
flux versus RM and the peak of the Gaussian was taken as the best RM value.  
The RM values quoted for each pulsar are average values over multiple observations with the
standard deviation as the uncertainty.  This was done in case the ionosphere contaminated any
RM measurements (see, e.g., Yan et al.~2011).

\subsection{Pulse Widths at 820 MHz}\label{w50}

Pulse widths at 50\% intensity, $W_{50}$, were derived either by fitting a Gaussian for 
single peak profiles or by measuring 
the width across the outer edge of the profile at 50\%  of the peak flux for multiple 
component profiles.  This was done for each epoch. The $W_{50}$ values for each epoch 
were averaged and the standard deviation was taken as the uncertainty.

\section{Discussion}\label{dis}

\begin{figure}
\includegraphics[scale=0.80,angle=270]{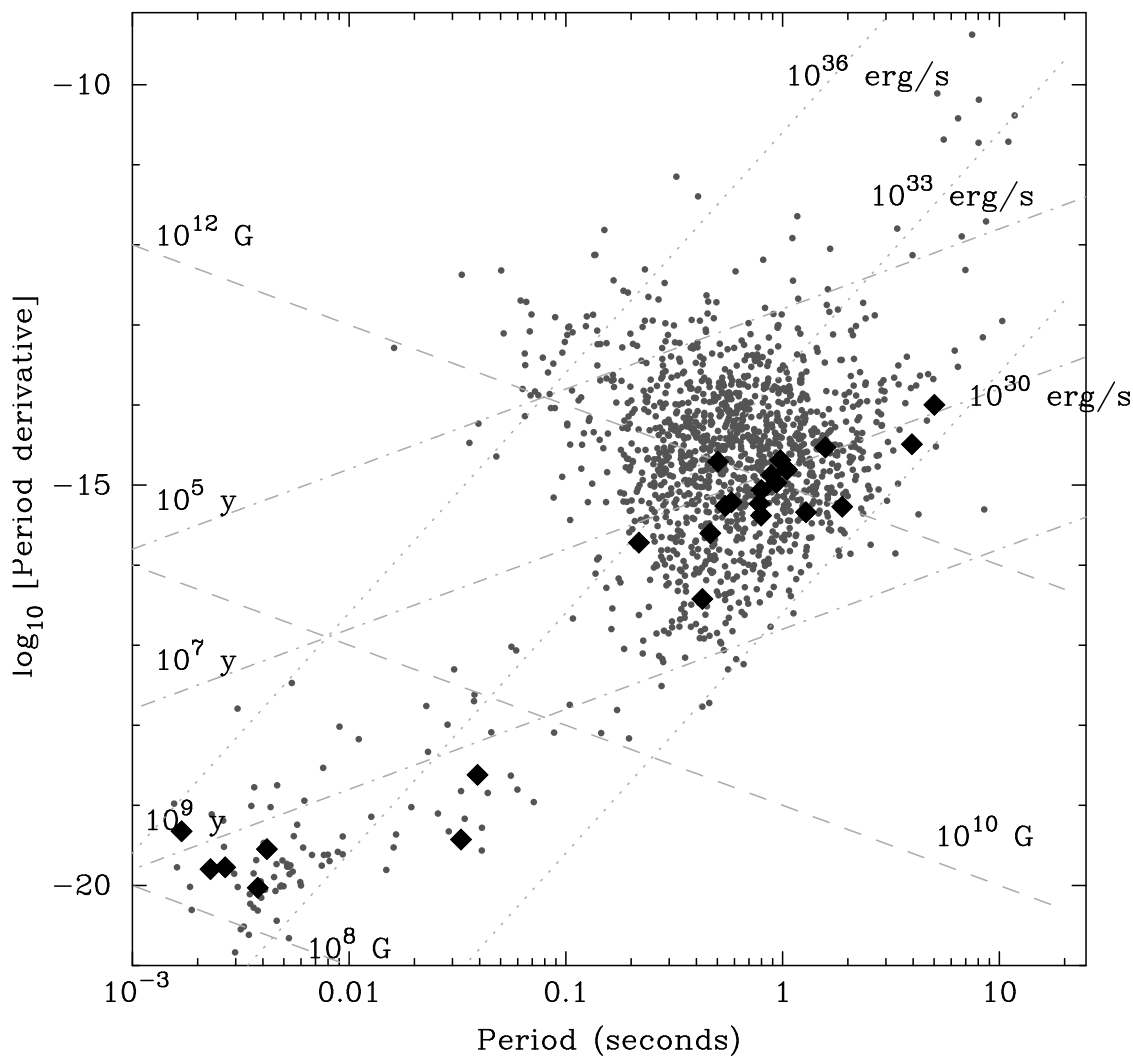}
\caption{\label{fg:ppdot}
A $P-\dot{P}$ diagram showing the new discoveries reported in this paper and 
the companion paper Paper II as 
black diamonds.  The lines on the plot show constant spin-down 
luminosity (dotted line), constant magnetic field (dash line), and constant 
characteristic age (dot-dash line).  The pulsar population is taken from the 
ATNF pulsar catalog (Manchester et al. 
2005, http://www.atnf.csiro.au/research/pulsar/psrcat/).} 
\end{figure}

Results from pulsar timing and the analysis described in Section~\ref{non} are located
in Tables 2 and 3 for each individual pulsar and the location 
of each new pulsar on the $P-\dot{P}$ diagram is given in Figure~\ref{fg:ppdot}. 
The DM-derived distances are from the NE2001 
electron density model (Cordes \& Lazio 2002).  The distances have uncertainties of approximately 
25\%.  The surface magnetic field strength, characteristic age, and spin-down 
luminosity are calculated using standard formulae (see, e.g. Lorimer \& Kramer 2005).   
Pulse profiles for each pulsar can be seen in Figures~6 and 7.  
Some of the individual pulsars are discussed in the following subsections.  

\subsection{PSR J1327$-$0755}

At the time of discovery, the narrow 350-MHz pulse width of 0.25 ms, period of 2.68 ms, and
location well off the Galactic plane ($b$ = 54.3$^\circ$)
of PSR J1327$-$0755 were encouraging signs that this
pulsar may be a good addition to pulsar timing arrays for the search for
gravitational waves (e.g., Foster \& Backer 1990).  However,  analysis at 820 MHz indicated that
PSR J1327$-$0755 may not meet the standards needed for a pulsar timing arrays pulsar due to 
very large variations in its flux from interstellar scintillation.  While $\sim$95$\%$ of GUPPI
observations resulted in a detection, only $\sim$35$\%$ of Spigot observations
resulted in a detection. The pulsar was typically detected in only
$\sim$25$\%$ of the 200-MHz GUPPI bandpass, due to interstellar scintillation.  
The orbit has a period of 8.44 days and an eccentricity so low it has not yet been detected 
in our timing: e = $9(6) \times 10^{-7}$.  The projected semi-major axis of 6.65 lt-s implies a 
minimum companion mass of 0.22 $\msun$, assuming a pulsar mass of 1.35 $\msun$. Given the pulsar's 
location in the $P-\dot{P}$ diagram, the companion is very likely to be a white dwarf.  

A figure of merit for the strong equivalence principle (SEP; Damour \& Schaefer 1991) test is 
the value of $\rm P_b^2/e$ (Lorimer \& Freire 2005).  This value for PSR~J1327$-$0755 is 
$\rm 7.91 \times 10^{7}~days^2$.  This values is within the range of values for other pulsars used 
for SEP tests and would be a good addition to arrays of pulsar used for SEP tests (Gonzalez et al. 2011).  

The composite proper motion for PSR~J1327$-$0755 is 99$\pm$23 
mas~$\rm yr^{-1}$.  This proper motion corresponds to a transverse velocity of 800 $\pm$ 200 km~$\rm s^{-1}$
for the predicted NE2001 electron distance of 1.7 kpc.  Only two recycled pulsars are known with larger
values for proper motion and they are PSR~J0437$-$4715 (Deller et al. 2009) and PSR~J1231$-$1411 
(Ransom et al. 2011).  Both of these pulsars are 
located within 0.5 kpc of the Earth.  The large proper motion most likely implies that the NE2001 model 
does not accurately predict distances off of the Galactic plane along this line of sight, or there are additional 
observational parameters that are not accounted for in PSR~J1327$-$0755's model.   

\subsection{PSR J1623$-$0841}

PSR J1623$-$0841 was discovered only in the single-pulse search with a period of 503 ms 
and was labeled a rotating radio transient (RRAT; McLaughlin et al. 2006).  Later observations at 350 MHz
with Spigot showed single pulses, but no detections were made at 820 MHz.  Gridding
observations were unable to localize the position better than the discovery position
because of its transient nature.  The final \texttt{TEMPO} position was 20$\arcmin$ away from the discovery
position (outside the GBT's 820-MHz beam), and a timing solution was only attainable
with the increased sensitivity of GUPPI and a dense set of observations at 350 MHz. A
full discussion of PSR J1623$-$0841 will be presented elsewhere (Boyles et al., in preparation).

\subsection{PSR J1737$-$0814} 
PSR~J1737$-$0814 is a 4.18 ms binary pulsar in a low-eccentric orbit ($e=0.000053$) 
with an orbital period of 79.38 days and a minimum companion mass of 0.06~$\msun$, assuming a pulsar mass 
of 1.35 $\msun$.  Phinney (1992) predicted a relationship between orbital period and eccentricity of 

\begin{equation}
<e^2>^{0.5} = 1.5 \times 10^{-4}(P_b/100~{\rm days})~{\rm for}~ P_b > P_{\rm crit} \approx 25~{\rm days},
\label{eq:pbvse}
\end{equation}

\noindent based on fluctuation dissipation theory in the second star.  
The orbital eccentricity for PSR~J1737$-$0814 
 is about a factor of two lower than predicted by the model, but within the model's
uncertainty.  This indicates that the system may have followed the expected 
binary evolution for longer-period recycled pulsars.  

In principle, the low eccentricity should make this pulsar an excellent candidate for
tests for violation of the Strong Equivalence Principle (Damour \& Schaefer 1991) as well as
the combined Lorentz-invariance and momentum-conservation strong-field
Parameterized Post-Newtonian parameter $\hat {\alpha_3}$ (Bell \& Damour 1996).
However, the companion mass predicted by the core-mass--orbital-period
relationship (Tauris \& Savonije 1999) is approximately 0.33\,$M_{\odot}$, meaning
that, unless the system inclination angle is about $15^{\circ}$ or less
(which has an {\it a priori} likelihood of less than 4\%), the system
may have had a different evolutionary history.  Since it is best to use a
set of sources that have followed the same evolutionary path for these tests
(Wex 2000, Gonzalez et al. 2011), PSR~J1737$-$0811 may not be suitable for inclusion
after all.  We note that PSR~J1327$-$0755 has a minimum companion mass very
close to the value predicted by the (Tauris \& Savonije 1999) relation, therefore it {\it is}
a pulsar that should be included in future editions of the ensemble tests
for these parameters that would indicate a departure from General Relativity.

\subsection{PSR J1941+0121}

\begin{figure}
\includegraphics[scale=0.75,angle=270]{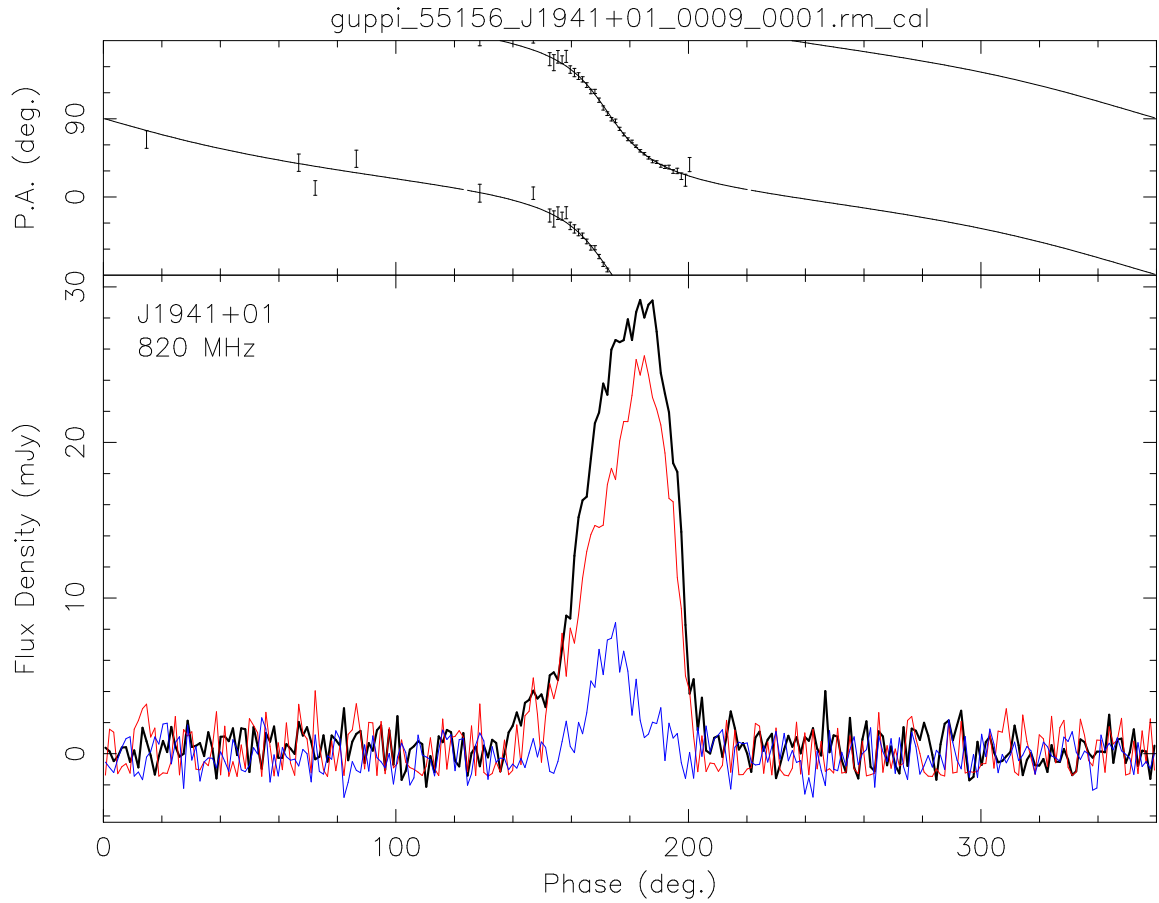} 
\caption{\label{fg:rmmod}
RVM model fit for the radio emission of PSR J1941+0121.  The top plot shows parallactic angle
versus pulse phase, while the bottom plot shows integrated pulse flux density versus pulse
phase for total intensity (black line), linearly polarized flux (red line), and circularly
polarized flux (blue line).}
\end{figure}

PSR J1941+0121 has the shortest period (217 ms) of the non-recycled pulsars found in this survey
and about 85\% of its total flux is linearly polarized.  This high percentage of linearly
polarized flux, along with its
relatively wide pulse width ($\epsilon$ = 0.093, where $\epsilon$ is the pulse width divided by pulse period), 
allowed us to produce a model 
of the pulsar's emission using the \texttt{PSRCHIVE} software package.  A complex-value Rotating
Vector Model (RVM; Radhakrishnan \& Cooke 1969) was fit to the Stokes Q and Stokes U
profiles, treating them as real and imaginary numbers instead of the single parameter
value of position angle.  Figure~\ref{fg:rmmod} shows the best fit of the model and
Table~1 gives the model fit parameters.  The model gives an impact parameter (i.e. angle between magnetic field 
axis and line of sight, $\beta$) of 8$^\circ \pm 4^\circ$ 
and a best-fit inclination angle (i.e. angle between rotation axis and magnetic field axis, 
$\alpha$) of 138$^\circ \pm 32^\circ$ 
These parameter values indicate an inner line of sight for this pulsar.  
The model has a $\chi^2$ of 23.2 with 31 degrees of freedom (on-pulse phase bins minus number of fitted parameters),
giving a reduced $\chi^2$ of 0.74.

\subsection{PSR J2222$-$0137}\label{sec:J2222}

PSR J2222$-$0137 is in a low-eccentricity orbit ($e=0.00038$)
with an orbital period of 2.4 days and a projected semi-major axis of
10.8~lt-s.  The orbital parameters imply a mass function of
0.229~$\msun$.  Assuming a pulsar mass of 1.35~$\msun$ results in a 
minimum companion mass of 1.11~$\msun$.  The spin period of 33~ms, the binary
parameters, and the lack of a large $\dot{P}$ indicate that
this pulsar is a partially recycled pulsar. Based on these parameters
and what is known about other binary pulsars, this object
is most likely a member of the so-called ``intermediate-mass binary
pulsars'' (e.g.~Camilo et al.~2001) in which the companion star
is a CO white dwarf. It is also possible, though less likely, given
the low orbital eccentricity,
that PSR J2222$-$0137 is a member of a double neutron star binary system.
About a dozen of the early follow-up observations of PSR J2222$-$0137 have
been searched for a second pulsar without success with a minimum detectable 
flux density of 11 $\mu$Jy at 1400 MHz.  

The presence of a second period derivative,
suggests that there are unmodeled effects in the residuals. 
An unmodeled Shapiro delay signal (as the pulsar's emission passes through the gravitational 
potential of the companion) and no measured proper motion could be contributing to the second period derivative.
We are not yet able to fit for this delay, but with
a minimum companion mass of 1.11 $\msun$ and narrow pulse duty cycle of 0.017,
it should be measurable, leading to the companion mass and orbital inclination. 
Due to the possible effects of unmodeled variables and a reduced $\chi^2$ of 13.3, 
no EFAC is included for PSR~J2222$-$0137's timing model, given in Table~3.  
An updated Shapiro delay and VLBA proper motion measurement,
along with an updated timing solution,  will be reported 
in a future publication.  

Using the DM value of 3.28 $\rm pc~cm^{-3}$, the NE2001 electron
density model predicts a distance of $\sim$300~pc to PSR J2222$-$0137
(Cordes \& Lazio 2002). If correct,
PSR J2222$-$0137 would be the second closest binary pulsar system after 
PSR J0437$-$4715, which has a DM of 
2.64 $\rm pc~cm^{-3}$ and a distance of 156~pc (Deller et al. 2009).
For pulsars located this close to us, a large
window of multi-wavelength observations opens up.  These will include VLBA
observations to measure proper motion and parallax, optical observations
to search for an optical companion, X-ray observations to look
for blackbody emission from the neutron star, and gamma-ray observations
to look for higher magnetosphere emission.  These observations will be reported in a
future publication.

\section{Conclusions and Future Work}\label{con}

The 350-MHz drift-scan survey revealed many interesting pulsars with a range of physical questions that 
they can address.  
Along with PSR J2222$-$0137, featured in
this work, PSR J1023+0038 (Archibald et al. 2009), PSR J2256$-$1024 (Stairs et al. in prep), and
PSR J0348+0438 (Paper II) have been studied in greater detail.  
Due to the success of this survey, a new survey named the Green Bank North Celestial Cap (GBNCC) 
survey is underway (Stovall et al. in prep).
In this survey we are taking advantage of the newest pulsar instrument on the GBT,
the GUPPI backend, which allows 100-MHz of bandwidth at 350 MHz.  
This survey will be 30$-$40\% more sensitive than the survey described in 
this paper and will be a powerful probe of the millisecond pulsar population in the
northern sky.  The GBNCC survey will eventually cover the whole entire northern sky visible by
the GBT.

\section*{Acknowledgments}

JB acknowledges support from WVEPSCoR, the National Radio 
Astronomy Observatory, the National Science Foundation (AST 0907967), and the 
Smithsonian Astrophysical Observatory (Chandra Proposal 12400736).  
R.S.L. was a student at the National Radio Astronomy Observatory and
was supported through the GBT Student Support program and the National
Science Foundation grant AST-0907967 during the course of this work.
J.W.T.H. is a Veni Fellow 
of the Netherlands Foundation for Scientific Research.  Pulsar research at UBC is supported 
by NSERC and the CFI.  V.M.K. holds the Lorne Trottier Chair in Astrophysics and Cosmology, and a
Canada Research Chair, a Killam Research Fellowship, and acknowledges additional support
from an NSERC Discovery Grant, from FQRNT via le Centre de Recherche Astrophysique
du Qu\`{e}bec and the Canadian Institute for Advanced Research.  
MAM and DRL are supported by WVEPSCOR, NSF PIRE award $\#0968296$, and the Research Corporation 
for Scientific Advancement.
Pulsar research at UBC is supported by an NSERC Discovery Grant and 
Special Research Opportunity grant as well as the Canada Foundation for Innovation.  JB 
would like to thank Paul Demorest for use of his {\tt psrzap} utility.  
JB would like to thank Paulo Freire for his post-submission comments and useful feedback.  
R.F.C., C.E.R., and T.P. were summer students at the National Radio 
Astronomy Observatory during a portion of this work.  
We are also grateful to NRAO for a grant that assisted data storage.  
The National Radio Astronomy Observatory is a facility of the National Science 
Foundation operated under cooperative agreement by Associated Universities, Inc.

\clearpage 

\begin{table}
\label{tb:mod}
\begin{center}
\begin{tabular}{lcc}
\hline
Parameter & Angle & Error  \\
 & (degrees) & (degrees)  \\
\hline
$\psi_0$ & $-79$ & 3 \\
$\beta$ & 8 & 4 \\
$\alpha$ & 138 & 32 \\
$\phi_0$ & 351.7 & 0.8 \\
\hline
\end{tabular}
\caption{Geometric beam model fit for PSR J1941+0121 using a complex-number RVM, where $\phi_0$ is
the pulse phase of magnetic meridian, $\psi_0$ is position angle at $\phi_0$, $\alpha$ is the 
colatitude of magnetic axis, and $\beta$ is the impact parameter.
 }
\end{center}
\end{table}

\begin{table}
\label{tb:par1}
\begin{center}
\begin{tabular}{lcc}
\hline
\hline
\multicolumn{1}{c}{Parameter} & PSR J1555$-$0515 & PSR J1612+2008 \\
\hline
 & &  \\
\hline
\multicolumn{2}{c}{Timing Parameters} \\
\hline
Right Ascension (J2000) & 15:55:40.097(16) & 16:12:23.432(3) \\
Declination (J2000) & $-$05:15:57.4(5) & 20:08:18.33(4) \\
Spin Period (s) & 0.97540989108(5) & 0.4266459810253(12) \\
Period Derivative ($\rm s~s^{-1}$) & 2.054(2)$\times 10^{-15}$ & 3.727(15)$\times 10^{-17}$ \\
Dispersion Measure (pc $\rm cm^{-3}$) & 23.46(3) & 19.544(10) \\
Reference Epoch (MJD) & 54985 & 54985 \\
Span of Timing Data & 54635--55335 & 54635--55335 \\
Number of TOAs & 36 & 41 \\
Weighted RMS Residual ($\mu$s) & 524 & 164 \\
EFAC & 3.22 & 1.72 \\
\hline
\multicolumn{2}{c}{Derived Parameters} \\
\hline
Galactic Longitude (degrees) & 3.97 & 35.52 \\
Galactic Latitude (degrees) & 34.97 & 43.74 \\
Distance (kpc) & 1.2 & 1.3 \\
Distance From Plane (kpc) & 0.69 & 0.90 \\
Surface Magnetic Field ($10^{12}$ Gauss) & 1.4 & 0.13 \\
Spin-down Luminosity ($\rm 10^{31}~ergs~s^{-1}$) & 8.8 & 1.9 \\
Characteristic Age (Myr) & 7.4 & 180 \\
820 MHz Flux Density (mJy) & 1.2(2) & 0.83(11) \\
$\rm W_{50}$ 820 MHz (ms) & 12.9(3) & 3.29(10) \\
Rotation Measure (rad $\rm m^{-2}$) & 1.3(3.0) & 22(3) \\
\hline
\multicolumn{3}{l}{
}
\end{tabular}
\caption{Timing and derived parameters for newly discovered isolated pulsars.
}
\end{center} 
\end{table} 

\addtocounter{table}{-1}
\begin{table*}
\label{tb:par2}
\begin{center}
\begin{tabular}{lcc} 
\hline
\hline
\multicolumn{1}{c}{Parameter} & PSR J1623$-$0841 & PSR J1633$-$2009 \\
\hline
 & &  \\
\hline
\multicolumn{2}{c}{Timing Parameters} \\
\hline
Right Ascension (J2000) & 16:23:42.701(19) & 16:33:55.30(6) \\
Declination (J2000) & $-$08:41:36.4(9) & $-$20:10:09(5) \\
Spin Period (s) & 0.503014997755(14) & 0.93555704483(6) \\
Period Derivative ($\rm s~s^{-1}$) & 1.958(3)$\times 10^{-15}$ & 1.070(3)$\times 10^{-15}$  \\
Dispersion Measure (pc $\rm cm^{-3}$) & 60.42(4) & 48.19(6) \\
Reference Epoch (MJD) & 55079 & 54993 \\
Span of Timing Data & 54635--55522 & 54651--55335 \\
Number of TOAs & 36 & 29 \\
Weighted RMS Residual ($\mu$s) & 670 & 633 \\
EFAC & 2.17 & 1.62 \\
\hline
\multicolumn{2}{c}{Derived Parameters} \\
\hline
Galactic Longitude (degrees) & 5.77  & 357.63 \\
Galactic Latitude (degrees) & 27.37 & 18.33  \\
Distance (kpc) & 3.3 & 1.6 \\
Distance From Plane (kpc) & 1.5 & 0.5 \\
Surface Magnetic Field ($10^{12}$ Gauss) & 1.0 & 1.0 \\  
Spin-down Luminosity ($\rm 10^{31}~ergs~s^{-1}$) & 63 & 5.0 \\
Characteristic Age (Myr) & 4.1 & 14 \\
820 MHz Flux Density (mJy) & 0.42(13) & 1.35(4) \\
$\rm W_{50}$ 820 MHz (ms) & 11.2(3) & 32.2(1.0) \\
Rotation Measure$^a$ (rad $\rm m^{-2}$) & $\cdots$ & $-$0.1(1.2) \\
\hline
\multicolumn{3}{l}{$^a$ PSR J1623$-$0841 does not have enough linearly polarized flux for a RM measurement.
} \\
\end{tabular}
\caption{Timing and derived parameters for newly discovered isolated pulsars. 
}
\end{center}
\end{table*}

\addtocounter{table}{-1}
\begin{table*}
\label{tb:par3}
\begin{center}
\begin{tabular}{lcc}
\hline
\hline
\multicolumn{1}{c}{Parameter} & PSR J1735$-$0243 & PSR J1903$-$0848 \\
\hline
 & &  \\
\hline
\multicolumn{2}{c}{Timing Parameters} \\
\hline
Right Ascension (J2000) & 17:35:48.1(2) & 19:03:11.271(18) \\
Declination (J2000) & $-$02:43:48(5) & $-$08:48:57.4(8) \\
Spin Period (s) & 0.7828869765(5) & 0.88732464056(4) \\
Period Derivative ($\rm s~s^{-1}$) & 6.32(12)$\times 10^{-16}$ & 1.330(2)$\times 10^{-15}$ \\
Dispersion Measure (pc $\rm cm^{-3}$) & 55.4(5) & 66.99(4) \\
Reference Epoch (MJD) & 54827 & 54987 \\
Span of Timing Data & 54352--55302 & 54634--55339 \\
Number of TOAs & 39 & 51 \\
Weighted RMS Residual ($\mu$s) & 6532 & 824 \\
EFAC & 3.70 & 3.32 \\
\hline
\multicolumn{2}{c}{Derived Parameters} \\
\hline
Galactic Longitude (degrees) & 21.56 & 26.38 \\
Galactic Latitude (degrees) & 15.45 & $-$6.60  \\
Distance (kpc) & 1.9 & 2.0 \\
Distance From Plane (kpc) & 0.51 & 0.23 \\
Surface Magnetic Field ($10^{12}$ Gauss) & 0.68 & 1.1 \\  
Spin-down Luminosity ($\rm 10^{31}~ergs~s^{-1}$) & 4.7 & 7.2 \\
Characteristic Age (Myr) & 21 & 11 \\
820 MHz Flux Density (mJy) & 1.3(3) & 1.4(2) \\
$\rm W_{50}$ 820 MHz (ms) & 102(2) & 12.3(2) \\
Rotation Measure (rad $\rm m^{-2}$) & 28.3(1.9) & 4.4(9) \\
\hline
\multicolumn{3}{l}{
}
\end{tabular}
\caption{Timing and derived parameters for newly discovered isolated pulsars.
}
\end{center}
\end{table*}

\addtocounter{table}{-1}
\begin{table*}
\label{tb:par4}
\begin{center}
\begin{tabular}{lcc}
\hline
\hline
\multicolumn{1}{c}{Parameter} & PSR J1941+0121 & PSR J2012$-$2029 \\
\hline
 & &  \\
\hline
\multicolumn{2}{c}{Timing Parameters} \\
\hline
Right Ascension (J2000) & 19:41:16.039(16) & 20:12:46.6(3) \\
Declination (J2000) & 01:21:39.5(5) & $-$20:29:31(22) \\
Spin Period (s) & 0.217317451828(10) & 0.54400187061(3) \\
Period Derivative ($\rm s~s^{-1}$) & 1.913(5)$\times 10^{-16}$ & 5.475(18)$\times 10^{-16}$ \\
Dispersion Measure (pc $\rm cm^{-3}$) & 52.7(7) & 37.67(8) \\
Reference Epoch (MJD) & 55026 & 54987 \\
Span of Timing Data & 54712--55339 & 54635--55339 \\
Number of TOAs & 33  & 33 \\
Weighted RMS Residual ($\mu$s) & 555 & 538 \\
EFAC & 1.98 & 1.59 \\
\hline
\multicolumn{2}{c}{Derived Parameters} \\
\hline
Galactic Longitude (degrees) & 39.91 & 22.41 \\
Galactic Latitude (degrees) & $-$10.43 & $-$26.70 \\
Distance (kpc) & 2.2 & 1.4 \\
Distance From Plane (kpc) & 0.40 & 0.63 \\
Surface Magnetic Field ($10^{12}$ Gauss) & 0.20 & 0.55 \\
Spin-down Luminosity ($\rm 10^{31}~ergs~s^{-1}$) & 72 & 14 \\
Characteristic Age (Myr) & 18 & 16 \\
820 MHz Flux Density (mJy) & 1.7(5) & 1.00(12) \\
$\rm W_{50}$ 820 MHz (ms) & 20.2(1.1) & 24.2(7) \\
Rotation Measure$^a$ (rad $\rm m^{-2}$) & $-$92.0(3) & $\cdots$ \\
\hline
\multicolumn{3}{l}{$^a$ PSR J2012$-$2029 does not have enough linearly polarized flux for a RM measurement.
} \\
\end{tabular}
\caption{Timing and derived parameters for newly discovered isolated pulsars.
}
\end{center}
\end{table*}

\addtocounter{table}{-1}
\begin{table*}
\label{tb:par5}
\begin{center}
\begin{tabular}{lcc}
\hline
\hline
\multicolumn{1}{c}{Parameter} & PSR J2033$-$1938 & PSR J2111+2106 \\
\hline
 & &  \\
\hline
\multicolumn{2}{c}{Timing Parameters} \\
\hline
Right Ascension (J2000) & 20:33:55.4(2) & 21:11:33.13(6) \\
Declination (J2000) & $-$19:38:59(12) & 21:06:07.0(1.4) \\
Spin Period (s) & 1.28171902379(12) & 3.9538529596(6) \\
Period Derivative ($\rm s~s^{-1}$) & 4.55(6)$\times 10^{-16}$ & 3.24(2)$\times 10^{-15}$ \\
Dispersion Measure (pc $\rm cm^{-3}$) & 23.47(9) & 59.74(14) \\
Reference Epoch (MJD) & 54987 & 54987 \\
Span of Timing Data & 54635--55339 & 54635--55339 \\
Number of TOAs & 37  & 35 \\
Weighted RMS Residual ($\mu$s) & 606 & 2106 \\
EFAC & 2.05 & 2.33 \\
\hline
\multicolumn{2}{c}{Derived Parameters} \\
\hline
Galactic Longitude (degrees) & 25.32 & 69.40  \\
Galactic Latitude (degrees) & $-$31.04 & $-$18.20 \\
Distance (kpc) & 1.00 & 3.8 \\
Distance From Plane (kpc) & 0.51 & 1.2 \\
Surface Magnetic Field ($10^{12}$ Gauss) & 0.76 & 3.6 \\
Spin-down Luminosity ($\rm 10^{31}~ergs~s^{-1}$) & 0.85 & 0.20 \\
Characteristic Age (Myr) & 44 & 19 \\
820 MHz Flux Density (mJy) & 1.13(15) & 1.08(10) \\
$\rm W_{50}$ 820 MHz (ms) & 32.4(3) & 45(2) \\
Rotation Measure (rad $\rm m^{-2}$) & $-$17.7(5) & $-$75.3(8) \\
\hline
\multicolumn{3}{l}{
}
\end{tabular}
\caption{Timing and derived parameters for newly discovered isolated pulsars.
}
\end{center}
\end{table*}

\begin{table*}
\label{tb:par6}
\footnotesize
\begin{center}
\begin{tabular}{lcc}
\hline
\hline
\multicolumn{1}{c}{Parameter} & PSR J1327$-$0755$^a$ & PSR J1737$-$0811 \\
\hline
 & &  \\
\hline
\multicolumn{2}{c}{Timing Parameters} \\
\hline
Right Ascension (J2000) & 13:27:57.5880(12) & 17:37:47.11235(14) \\
Declination (J2000) & $-$07:55:29.80(4) & $-$08:11:08.887(5) \\
RA Proper Motion & 27(18) & $\cdots$ \\
DEC Proper Motion & 95(48) & $\cdots$  \\
Spin Period (s) & 0.0026779231971205(2) & 0.0041750173128551(8) \\
Period Derivative$^b$ ($\rm s~s^{-1}$) & 1.773(18)$\times 10^{-20}$ & 7.93(8)$\times 10^{-21}$ \\
Dispersion Measure (pc $\rm cm^{-3}$) & 27.91215(6) & 55.311(3) \\
Reference Epoch (MJD) & 55131 & 54987 \\
Span of Timing Data & 54649--55613 & 54635--55339 \\
Number of TOAs$^c$ & 188  & 225 \\
Weighted RMS Residual ($\mu$s) & 5.5 & 14 \\
EFAC & 1.33 & 1.08 \\
\hline
\multicolumn{2}{c}{Binary Parameters} \\
\hline
Binary Model & ELL1 & BT \\
Orbital Period (days) & 8.439086019(12) & 79.517379(2) \\
Projected Semi-major Axis (lt-s) & 6.645774(2) & 9.332791(3) \\
Time of Ascending Node (MJD) & 54717.3830798(7)   &  $\cdots$ \\
First Laplace-Lagrange Parameter & $6(5) \times 10 ^{-7}$ &  $\cdots$ \\
Second Laplace-Lagrange Parameter & $6(7) \times 10 ^{-7}$  &  $\cdots$ \\
Epoch Of Periastron (MJD) & 54718.50769778(9$\times 10^{-1}$) & 54696.879781933(2$\times 10^{-1}$) \\
Orbital Eccentricity & 9(6)$\times 10^{-7}$ & 5.38(8)$\times 10^{-5}$ \\
Longitude of Periastron (degrees) & 48(40) & 49.8(9) \\
Mass Function ($\msun$) & 0.004425158(4) & 0.0001380362(2) \\
\hline
\multicolumn{2}{c}{Derived Parameters} \\
\hline
Minimum Companion Mass$^d$ ($\msun$)& 0.2219585(9) & 0.0662357(3) \\
Galactic Longitude (degrees) & 318.39 & 16.86 \\
Galactic Latitude (degrees) & 53.85 & 12.34 \\
Distance (kpc) & 1.7 & 1.7 \\
Distance From Plane (kpc) & 1.0 & 0.36 \\
Surface Magnetic Field$^b$ ($10^{9}$ Gauss) & 0.70 & 1.8 \\
Spin-down Luminosity$^b$ ($\rm 10^{31}~ergs~s^{-1}$) & 360 & 430 \\
Characteristic Age$^b$ (Gyr) & 2.4 & 8.3 \\
820 MHz Flux Density (mJy) & 0.9(2) & 2.7(2) \\
$\rm W_{50}$ 820 MHz (ms) & 0.131(5) & 0.52(6) \\
Rotation Measure$^e$ (rad $\rm m^{-2}$) & $\cdots$ & 71.4(1.4) \\
\hline
\multicolumn{3}{l}{$^a$ Epoch Of Periastron, Orbital Eccentricity, and Longitude of Periastron are derived} \\
\multicolumn{3}{l}{values for PSR~1327$-$0755.  } \\
\multicolumn{3}{l}{$^b$ Values are not corrected for Shklovskii effect (Shklovskii 1970).} \\
\multicolumn{3}{l}{$^c$ Three TOAs produced per observation.} \\
\multicolumn{3}{l}{$^d$ Assumes a pulsar mass of 1.35$\msun$.} \\
\multicolumn{3}{l}{$^e$ PSR J1327$-$0755 does not have enough linearly polarized flux for a RM measurement.} \\

\end{tabular}
\caption{Timing, binary, and derived parameters for newly discovered binary pulsars.
}
\end{center}
\end{table*}

\addtocounter{table}{-1}
\begin{table}
\label{tb:par7}
\begin{center}
\begin{tabular}{lc}
\hline
\hline
\multicolumn{1}{c}{Parameter} & PSR J2222$-$0137 \\
\hline
 & \\
\hline
\multicolumn{2}{c}{Timing Parameters} \\
\hline
Right Ascension (J2000) & 22:22:05.96345(8) \\
Declination (J2000) & $-$01:37:15.627(3) \\
Spin Period (s) & 0.032817859050650(5) \\
Period Derivative ($\rm s~s^{-1}$) & 4.74(3)$\times 10^{-20}$ \\
Second Period Derivative ($\rm s~s^{-2}$) & 1.61(8)$\times 10^{-27}$ \\ 
Dispersion Measure (pc $\rm cm^{-3}$) & 3.27511(10) \\
Reference Epoch (MJD) & 55285 \\
Span of Timing Data & 54985--55585 \\
Number of TOAs$^a$ & 219 \\
Weighted RMS Residual ($\mu$s) & 8.2 \\
\hline
\multicolumn{2}{c}{Binary Parameters} \\
\hline
Binary Model & BT \\
Orbital Period (days) & 2.4457599960(5) \\
Projected Semi-major Axis (lt-s) & 10.8480469(8) \\
Orbital Eccentricity &  0.00038454(14) \\
Longitude of Periastron (degrees)  & 119.863(14) \\
Time of periastron (MJD) & 55284.775883844($1.0 \times 10^{-4}$) \\
Mass Function ($\msun$) & 0.22914466(5) \\
\hline
\multicolumn{2}{c}{Derived Parameters} \\
\hline
Minimum Companion Mass$^b$ ($\msun$) & 1.1169752(3) \\
Galactic Longitude (degrees) & 62.02 \\
Galactic Latitude (degrees) & $-$46.08 \\
Distance (kpc) & 0.31 \\
Distance From Plane (kpc) & 0.22 \\
Surface Magnetic Field ($10^{9}$ Gauss) & 1.3 \\
Spin-down Luminosity ($\rm 10^{31}~ergs~s^{-1}$) & 5.3 \\
Characteristic Age (Gyr) & 11 \\
820 MHz Flux Density (mJy) & 2.6(5) \\
$\rm W_{50}$ 820 MHz (ms) & 0.570(5) \\
Rotation Measure (rad $\rm m^{-2}$) & 1.8(6) \\
\hline
\multicolumn{2}{l}{$^a$ Three TOAs produced per observation.} \\
\multicolumn{2}{l}{$^b$ Assumes a pulsar mass of 1.35$\msun$.} \\
\end{tabular}
\caption{Timing, binary, and derived parameters for newly discovered PSR J2222$-$0137.
}
\end{center}
\end{table}

\begin{figure}[t]
\includegraphics[scale=0.8]{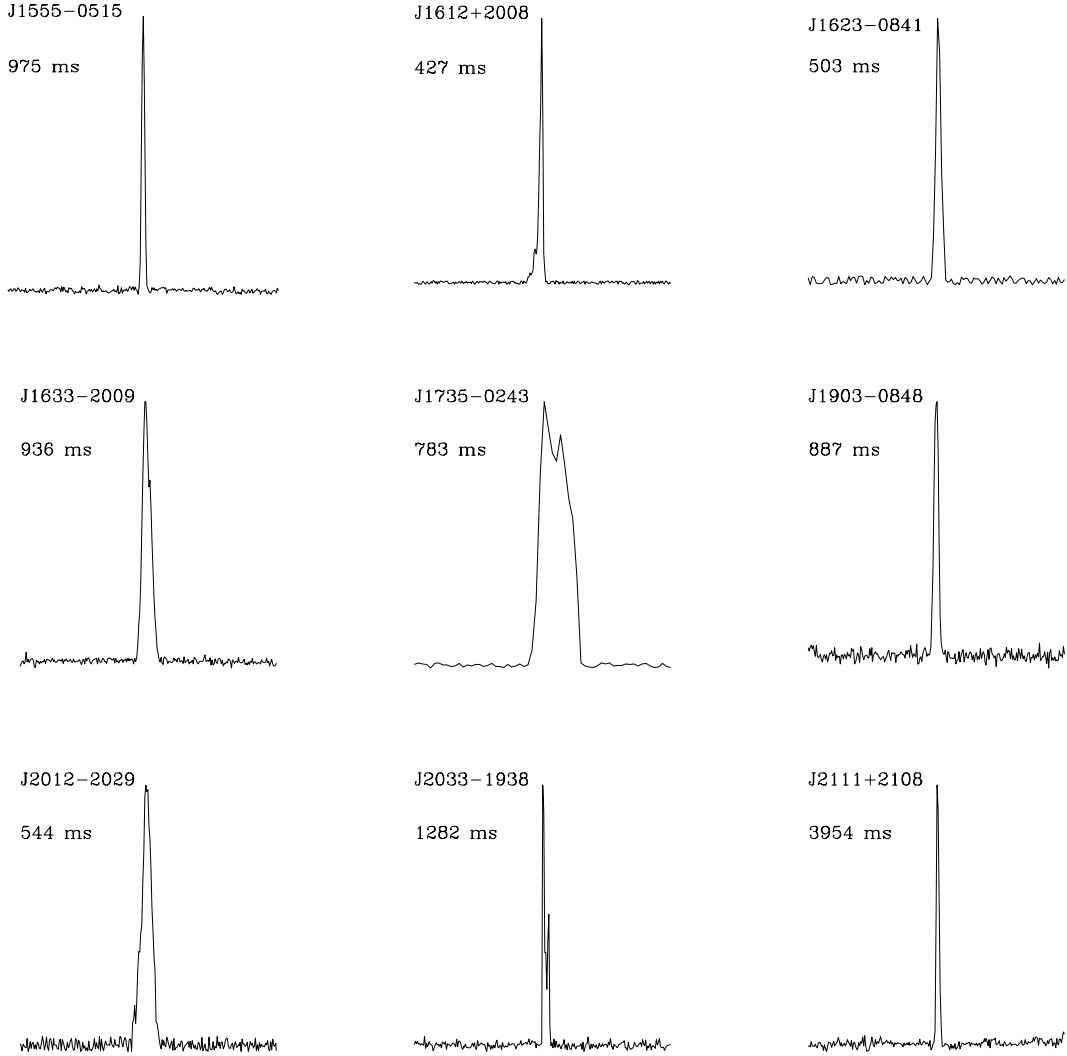}
\caption{\label{fg:prof}
Profiles at 820-MHz for nine newly discovered pulsars created from multi-epoch data using the GUPPI backend.  
Spin periods are listed.}
\end{figure}

\begin{figure}[t]
\includegraphics[scale=0.8]{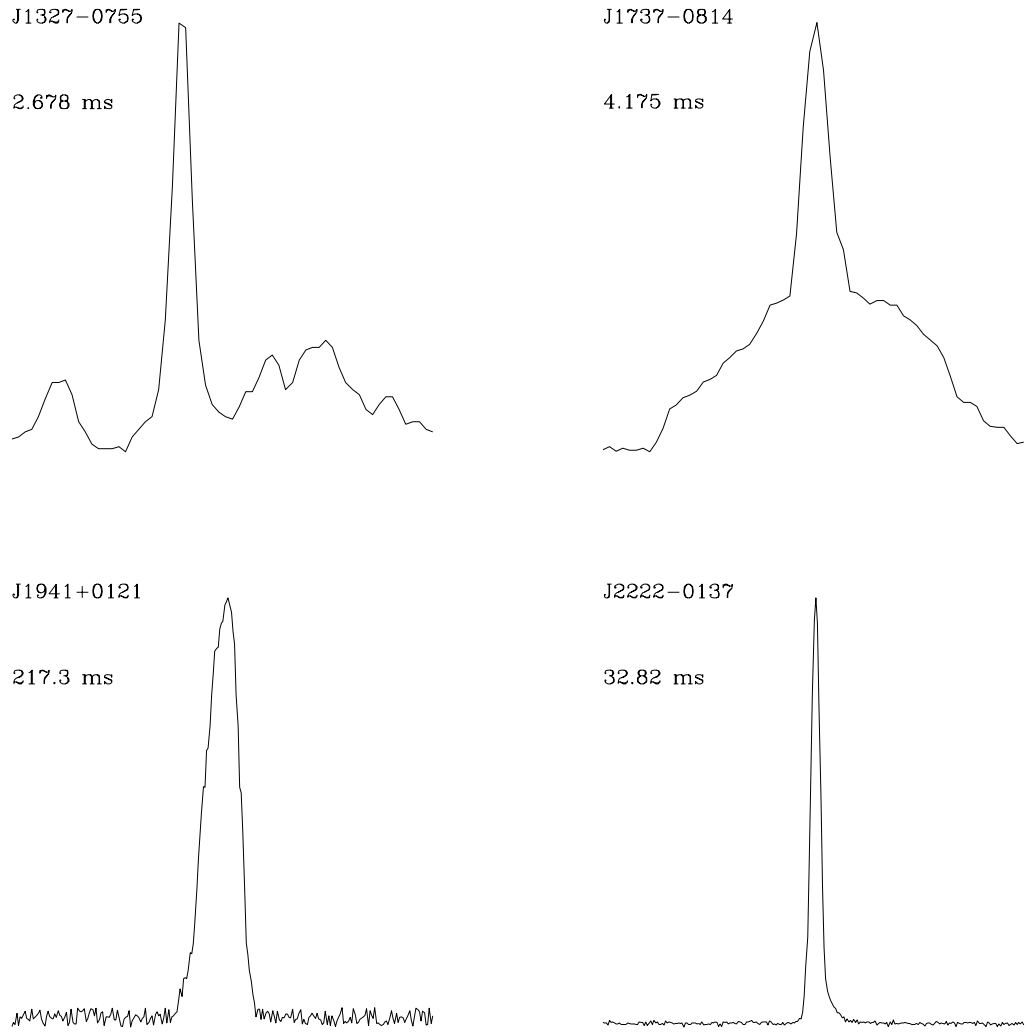}
\caption{\label{fg:msprof}
Profiles at 820-MHz for the remaining four newly discovered pulsars created from multi-epoch data using 
the GUPPI backend.  The spin periods are listed.}
\end{figure}

\end{document}